# PERFORMABILITY ASPECTS OF THE ATLAS VO; USING LMBENCH SUITE

Fotis Georgatos
*Department of Computer Science, University of Cyprus, Cyprus*
fotis@mail.cern.ch

John Kouvakis
*Department of Mathematics, University of the Aegean, Greece*
gkouvakis@hep.ntua.gr

John Kouretis
*Department of Physics, National Technical University of Athens, Greece*
ge01275@mail.ntua.gr

**Abstract**    The ATLAS Virtual Organization is grid's largest Virtual Organization which is currently in full production stage. Hereby a case is being made that a user working within that VO is going to face a wide spectrum of different systems, whose heterogeneity is enough to count as "orders of magnitude" according to a number of metrics; including integer/float operations, memory throughput (STREAM) and communication latencies. Furthermore, the spread of performance does not appear to follow any known distribution pattern, which is demonstrated in graphs produced during May 2007 measurements.

It is implied that the current practice where either "all-WNs-are-equal" or, the alternative of SPEC-based rating used by LCG/EGEE is an oversimplification which is inappropriate and expensive from an operational point of view, therefore new techniques are needed for optimal grid resources allocation.

## 1.    Acknowledgements

We hereby wish to thank our respective supervising professors, M. Dikaiakos, I. Gkialas, T. Alexopoulos as well as Dr. G. Tsouloupas for their support and constructive feedback at various opportunities during this work.

## 2. Introduction and outline

Grid computing emphasizes on sharing of heterogeneous resources, which might be based on different hardware or software architectures. In correspondence with this diversity of the infrastructure, the execution time of any single job, as well as the total grid performance can both be affected substantially. The objective of the current work is to document this effect with real-world data.

A microbenchmarking technique using lmbench [4]has been applied in order to explore if this assumption is true within the context of the ATLAS Virtual Organization; Indeed, results show that performance can vary, up to an order a magnitude and the effects are even more apparent within this larger VO, than what we found during earlier work within SEE VO [8][9].

Real grid characteristics serve as a proof that metrics-guided resource selection is nearly imperative, if not to optimally select resources, at least to specifically avoid ones which are known a priori that they don't perform as good as required. Our results hint in favour of a more intelligent matchmaking process which involves performance metrics.

## 3. Related Work

A similar approach was taken by the developers of the GridBench platform [1], which is a tool for evaluating the performance of Grids and Grid resources through benchmarking. Measurements were taken from CrossGrid and the LCG testbed for the purpose of resource characterization. Lately more tests have been done on the EGEE infrastructure with more conclusive results [2]. Indeed, such benchmarking techniques have already been demonstrated to be of interest [3]. The current static Information System-based practices should be augmented by dynamic characterization of grid resources. There exists evidence that performance can be improved for a few application categories, by applying metrics-correlation techniques on a case-by-case basis [1].

## 4. Issues and Methodology

Benchmarks are standardized programs or detailed specifications of algorithms designed to investigate well-defined performance properties of computer systems in alignment with a widely demonstrated set of methods and procedures. For many years, benchmarking has been used to characterize a large variety of systems ranging from CPU architectures and caches to filesystems, databases, parallel systems, Internet infrastructures and middleware. Computer benchmarking provides a commonly accepted basis for comparing the performance of different computer systems in a highly repeatable and consistent manner, so it appears appealing to use it for providing resource metrics.

LCG/EGEE grid currently provides insufficient information about sites' characteristics, which results in longer queue and job execution times and, indirectly, to more failures. The data available in the Information System is total memory of a node, Operating System distribution name and version, processor model and total CPUs per site. This information is not always complete and the commonly used dataset that provides sites' technical specifications, such as processor model, total memory and total cpus per site are often inaccurate or inconsistent, since they are manually edited, turning it far from optimal choices as selection criteria, while finding the most appropriate ones among sites.

The basic job submission framework of this endeavour is already available as an open source python code package, which is able to submit self-compiling lmbench sources along with the related scripts that gather other system information -software & hardware- and collect back their reports. It is available along with the rest of the code at: http://GetNRunBench.sourceforge.net

What is very important to specifically clarify, is that the benefits of applying the benchmarking technique and grid resource characterization can greatly outnumber the measurement system's overhead in itself; typically it can be run at a rate of 1 test/site/day consuming less than 0.5% of the ATLAS Virtual Organization capacity, traded for a benefit which can be much larger, or at least this is what we have conjectured from these first results.

## 5. Results

The third release of lmbench (lmbench-3.0-a4) has been selected for its better benchmarking techniques and more detailed and documented micro-benchmarking tools [5]. During the execution of this activity a Technical Report has been produced which includes the full set of processed results [7].

In parallel to the lmbench activity scripts were run which collected data on the grid environment, gathering information like: linux distribution and kernel, memory and swap configuration, CPU Model and Vendor as well as grid middleware. This information, once collected, can both document the heterogeneity of the grid, as well as provide input for processes that have particular needs for their execution environment. Please refer to the Technical Report for further details or the raw dataset available in the grid manner at 'lfn:/grid/see/fotis/ATLAS_lmbench_2007_Taurus_dataset_reduced-by-fotis.tgz'. We hereby provide a subset of the data in pie and histogram graphs.

The SizeOfSite distribution graph Figure 1 presents the amount of sites that exist within a given "size category". Sites range in size, and their capacity can be from a few jobs up to thousands. It is worthy to notice that the large majority of resources, about 60% is provided by the top 10% of sites. Also: about 75% by the top 20%, a 90% by just 40% of sites, 95% by the top 60% and 99% by the top 80% of sites. The last 20% of sites contribute just 1% of resources.

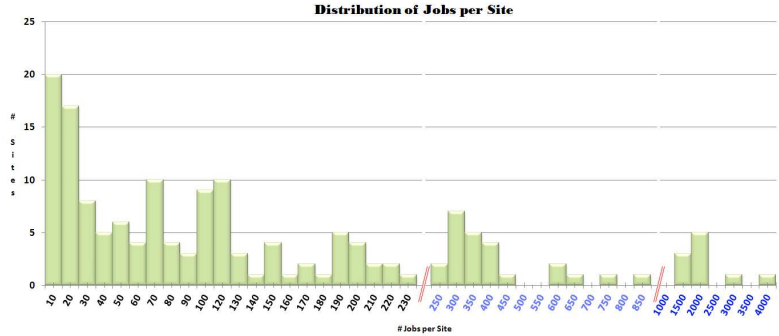

*Figure 1.*    SizeOfSite Distribution

In order to better visualize results and instead of providing only minima and maxima, we employed a "projection" technique: we attempted to derive histograms from our measurements combining data from the Information System (BDII & GSTAT), attempting to correlate our multiple measurements and site size, "pro-rata". Projection was unavoidable since benchmarking all of EGEE is impossible; The resulting graphs should be considered reliable on the horizontal axis, since all measurements have indeed taken place within the ATLAS VO, but with a grain of salt on the vertical axis since large sites can contribute a larger error margin. For the conclusions of this work, this effect is irrelevant; We are looking into ways of minimizing the error margin or find bounds for it, though.

As show in Figure 4, the most popular linux distribution is Scientific Linux CERN v3.0.8, which can be explained due to the endorsement by the LHC Computing Grid project: all large Tier-1 sites (many 100s or 1000s of CPUs) and Tier-2 sites have to use it: Any system administrator of an LCG site has to update her site promptly, in order to remain compatible with CERN experiments' software. The next most popular distribution is Scientific Linux v3.0.x. It is obvious that their dispersion is much higher, perhaps a by-product of their system administrators following a more conservative upgrade policy, combined with the lack of a centrally coordinated driving force, or just being driven by more diverse needs.

The amount of Linux distributions that are concurrently in use on the grid is extraordinary and include versions of CentOS, RedHat, SL, SLC, Ubuntu, Debian and some more customized RedHat clones; it should be considered fortunate if there are no applications failing due to that very reason: there are many software tools that expect to find software in specific path locations, esp. scripting languages like perl, python or php, or system libraries, so this variety of systems could very well increase job failure rates.

*Figure 2.* Environmentals

*Figure 3.* Linux Kernel and CPU Model

*Figure 4.* Linux Distributions

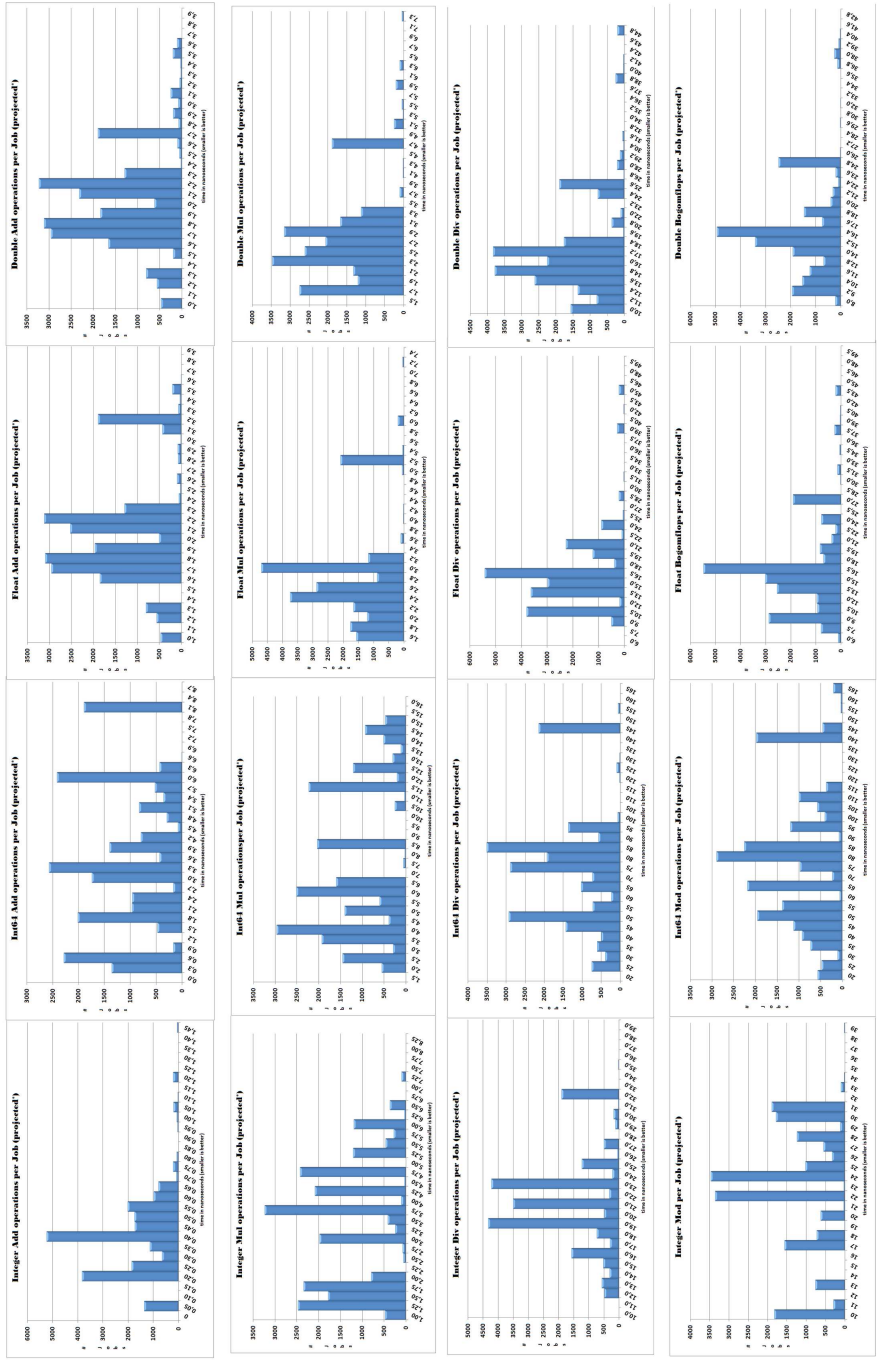

*Figure 5.* LMbench measurements - Numerical operations

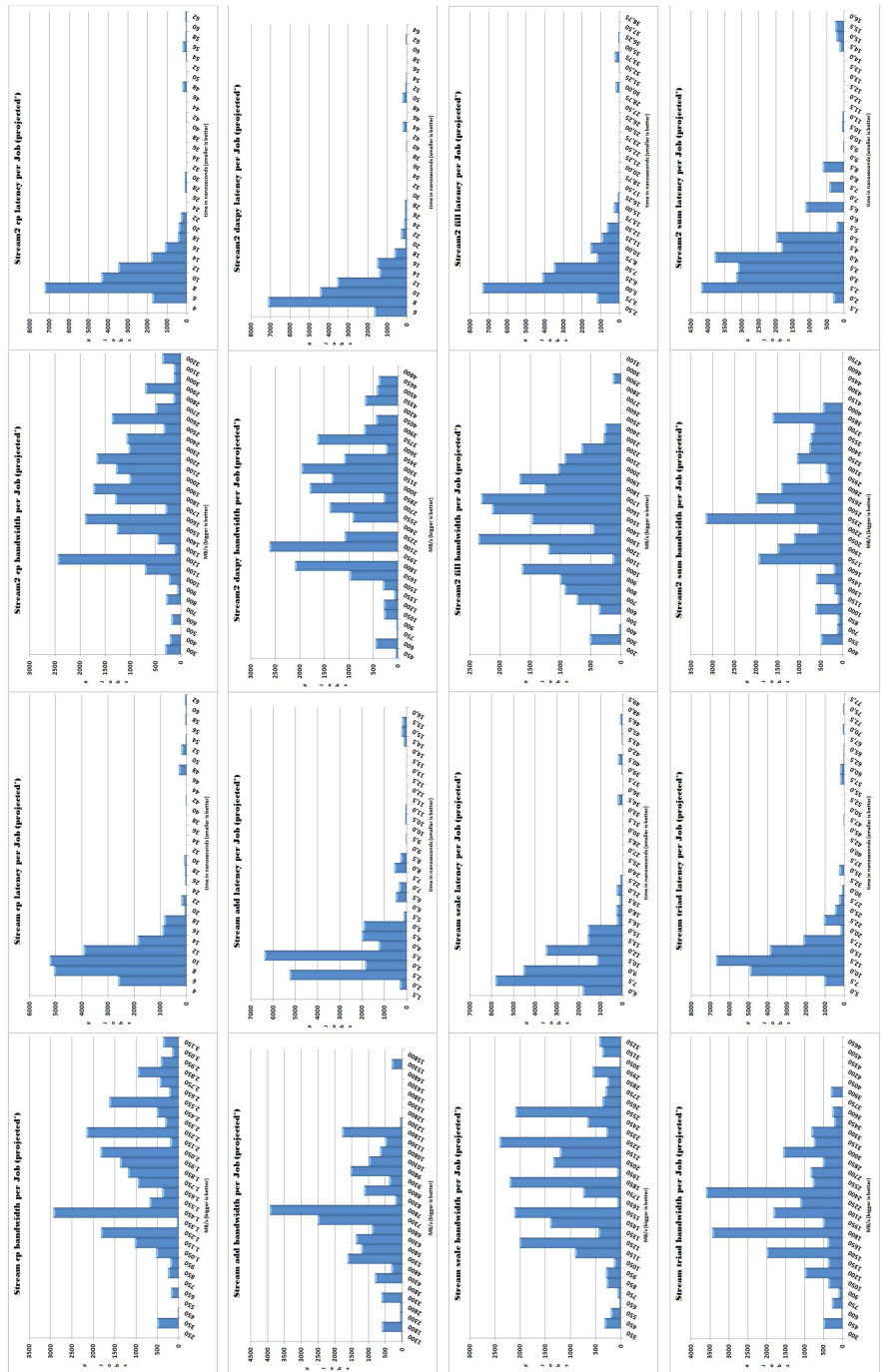

*Figure 6.* LMbench measurements - Stream, Stream2

It is worthy to note that certain benchmark histograms manifest interesting patterns. It is unlikely that they are poisson processes, at least as can be seen from a grid user point of view:

- Multiplication of 64bit integers "int64 mul" is a great example of grid performance effects: the measurements can be split into two areas, those of "fast" and "slow" clusters, thereby allowing us to apply during job submission two distinct strategies: optimize for throughput (use both sets of resources) or latency (use only fast ones). The values cluster between 1.5-8.5 nanoseconds and 10.0-15.0 nanoseconds; preliminary investigation hints that the effect is caused by architectural aspects, namely the support for EM64T/x86-64 operations, which were introduced during year 2004 for 32bit processors.

- Multiplication of simple precision floating point numbers (Float Mul) excibits a cluster effect in the area 1.5-3.5 nanoseconds, while there are peaks of performability at 5.2, 6.0 and 7.2 nanoseconds. For example, if we execute a DAG workflow on the grid for which we want to minimize the makespan, it could prove beneficial if we exclude resources with Float-Mul metric >3.5nanoseconds, trading-off a minor part of resources, assuming a float multiplication-dominant operation. Similar patterns and conclusions could be seen with double precision multiplication (Double Mul) operation.

- Addition of 64bit Integers is an operation that appears to happen in a pattern which does not follow any known distribution, between 0.3-8.1 nanoseconds. This in effect is a ratio greater than 1:25, which implies that the execution of any relevant workflow on the grid could easily end-up being last-job-bound instead of CPU-bound: the processes of a certain group would have to wait before the last one of them finishes, and this is quite easy to happen if this kind of operation is dominant. Excluding resources would be a difficult decision as well, because there is no "optimal sweetspot", so there is no obvious strategy, optimal strategy has to be considered on a case-by-case basis.

- As seen in Figure 7, doing two BIT operations of 32 bit data is apparently not the same as doing one 64 bit operation; in fact, the latter is slower, which is perhaps counter-intuitive. This kind of information can help grid algorithm developers to take the optimal decisions while constructing software to be run on the grid, at least in an instance of the grid as it has right now. The idea can be generalized further in other similar dilemmas, eg. when to trade a division operation for modulo and rotate operations etc.

*Figure 7.* Bit operations

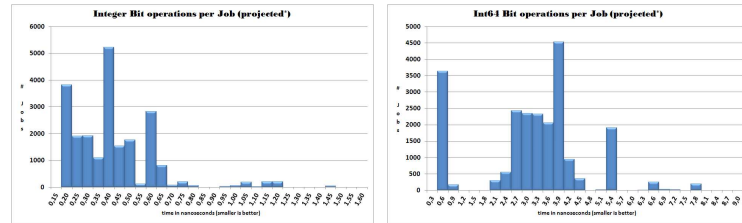

(a) Integer 32 bits  (b) Integer 64 bits

## 6. Analysis and Discussion

Timings of the various microbenchmarks are mostly being presented in nanoseconds. Even though they appear very small, according to these measurements, if they occupy a repetitive part of a grid job they can have an important and highly-impacting factor: A job that could be executed in a site in a time period T, in some other site could be executed in double that time, T*2; ignoring at this point any communication overheads. In fact, it is possible to find on the graphs ratios that are as high as 1:10, All STREAM benchmarks show this effect quite well, while the same holds true in particular with multiplication operations, or the operations involving many bits (more than 32, 64 bit operations plus simple and double floating point).

It is important to realize that extended job lengths can result in higher job failure rates, due to time window limits in sites' queues, which is standard practice.

In the special case that a grid job can be further parallelized in multiple components, each part could be sent in the most suitable site depending on the nature of the subprocess, and then the time of individual subprocesses will also be decreased, and consequently the total execution time as well (we assume no I/O overheads).

Finally, it is impossible for each user to know or measure the characteristics of each site. Therefore some mechanism must exist that allows the matchmaking to happen in an automatic way. There is some ongoing discussion if the best way to implement this would be through a job description technique (ie. in the .jdl file), or at the global scheduling stage (ie. RB or WMS), or both. The latter of course is advantageous, if it is combined with an Information System that can provide such benchmarking results; then it becomes possible for the middleware to identify the sites that are best for a specific job, assuming all other issues equal. For one thing, resource ranking is deemed necessary.

## 7. Future work

In order for resource ranking to be efficient, we have to remove the heterogeneity at the queue interface which is the optimal, most fine-grained, resource handle. Having multiple queues within a site, CE in gLite parlance, should be considered required if we are to support differentiated "subclusters", where Worker Nodes of similar performance aspects are grouped together. It is contemplated that if these metrics are consistent over time, it would be possible to optimize grid performance further, simply by profiling the current infrastructure, storing the results, then using submission frameworks that can understand this information. Alternatively, we could create and store ensembles of performance models [6], so that we are able to capture heterogeneity behind submission queues, and trigger mechanisms which provide for a "heterogeneity" metric: this will allow the user to trade bulk performance and uncertainty level for latency.

We hope that this information will be used for further cluster performance research and that it will help future system administrators choose better hardware and/or software components during the deployment of new clusters. In fact, a new era begins where instead of "brute-force" usage of resources, we will be able to load-balance grids according to their true capabilities, and purchase new Grid hardware upgrades according to real user needs, just as is envisaged in power, transportation and communication systems. For instance, should we buy more 64bit or, 32bit CPUs to upgrade a VO optimally?

We easily concluded during the analysis of raw data, that there areas for improvement, since this study is unusually sensitive to various deficiencies or shortcomings of data as they are provided by the Information System:

- heizenbugs; transient faulty data on the Information System

- dubious entries; permanent faulty data on the Information System

- duplicate Computing Elements or plainly wrong number of CPUs; there is difficulty in automatically extracting "safe" information

- lack of SubClusters feature support; by far the most serious issue, and the one for which we can do little about, yet.

Note also that once someone has to cope with a very high failure rate reaching nearly 50%, as we had, she will have to face overheads and unavoidably narrower findings. As said earlier though, the robustness of the technique is already severely limited by the native capabilities of the Information System itself, at least as we know its implementation right now with gLite-based grids (May 2007)

| | |
|---|---|
| + | better understanding of performance-related job-failures |
| + | better job scheduling with **substancial** throughput gains |
| + | capability to deliver prompt service for urgent computing |
| + | archival of performance aspects of the systems |
| - | system deployment overhead; but submission framework is ready |
| - | consumption of resources; can be optimized for continuous use |
| - | the technique is sensitive in resource heterogeneity (queue level) |

*Table 1.* Continuous Grid Benchmarking: The arguments of a debate

As a final point in this discussion, we would to report some arguments for and against **continuous grid benchmarking**, which are shown hereby.

## 8. Conclusions

The tool used in this report has been lmbench, due to its rich set of micro-benchmarks, as well as its availability and compatibility with Linux. It includes latency, bandwidth and timing measurements. Furthermore, information has been retrieved and reported about sites characteristics, such as linux kernel version and distribution, grid middleware, cpu and memory size.

The value of this work is to provide some initial real-world experimental data, in order to be able to evaluate possible strategies for future implementations.

The preliminary conclusion is that any typical grid can largely benefit from even trivial resource characterization and match-making techniques, assuming advantage is taken of this information early during job scheduling. In other words we could obtain knowledge about sites' performability with very little overhead and then take advantage of this information during scheduling to minimize total execution time considerably.

One more conclusion is the internal heterogeneity of large clusters is alone enough to drive much of the grid away from optimal performance: Once we can't differentiate among fast and slow Worker Nodes, we will be forced to cope with the uncertainty of makespan for any single job. This can seriously affect low-latency jobs which are characteristic of Urgent Computing applications, eg. earthquake and other natural hazard signal analysis, and impact badly the total workflow execution time, even prevent us from certain categories of domains, eg. a hurricane or Tsunami early warning system. The problem can be corrected, if the "subclusters" concept is put to use: in fact, we claim that if every single queue is always addressing a homogeneous resource pool, then we would be able to do queue-based characterization of availability, performance, stability; many aspects of the grid as we know it would be vastly improved.

We wish at this point to summarize our own understanding of the results:

a) Resource characterization and metrics driven-scheduling is imperative, if we seek optimal grid latency & throughput within a VO like ATLAS

b) Resource characterization is effective, as long as collected metrics correspond to true performance of grid sites: "heterogeneity brings uncertainty" and "uncertainty is expensive", so "heterogeneity is expensive",

Our results are quite conclusive in the aforementioned directions. We are eager to compare the results with those of other teams and make a bold statement on the need for resource auditing and ranking within the Information System itself. We claim that such an option can assist any Grid Infrastructure tremendously both in terms of throughput and latency and it is an implied requirement for any generic large scale grid system, if efficiency is actively seeked for.